\documentstyle[epsf]{mn}
\def\mmm{(m-M)$_0$}
\def\ebv{E(B$-$V)}

\def\bv{B$-$V}
\def\gsim{\;\lower.6ex\hbox{$\sim$}\kern-7.75pt\raise.65ex\hbox{$>$}\;}
\def\lsim{\;\lower.6ex\hbox{$\sim$}\kern-7.75pt\raise.65ex\hbox{$<$}\;}

\title[NGC 2506]{Old open clusters: UBGVRI photometry of 
 NGC 2506\thanks{Based on observations made at CTIO 
 and Mt. Palomar telescopes.
 CTIO is operated by the Association of Universities for Research in Astronomy 
 under agreement with the National Science Foundation.}}

\author[Marconi et al.]{G. Marconi$^1$, D. Hamilton$^2$, 
       M. Tosi$^3$, A. Bragaglia$^3$ \\
 $^1$ Osservatorio Astronomico di Monte Porzio, Italy, 
      e-mail marconi@coma.mporzio.astro.it\\
 $^2$ Institut f\"ur Astronomie $\&$ Astrophysik, Universit\"at M\"unchen, M\"unchen, Germany,
      e-mail hamilton@mpia-hd.mpg.de\\
 $^3$ Osservatorio Astronomico di Bologna, Italy, 
      e-mail angela@astbo3.bo.astro.it, tosi@astbo3.bo.astro.it}

\date{}

\begin{document}
\maketitle

\begin{abstract}
 UBGVRI photometry for the open cluster NGC 2506 is presented. From
 comparison of the observed colour-magnitude diagrams with simulations
 based on stellar evolutionary models we derive in a self consistent
 way reddening, distance, and age of the cluster:  \ebv =0-0.07, 
 (m-M)$_0$ = 12.6$\pm$0.1, $\tau$ = 1.5-2.2 Gyr. 
 The cluster shows a well definite secondary sequence, suggesting that 
 binary systems constitute $\gsim$ 20~\% of the cluster members visible
 in the colour-magnitude diagram.
\end{abstract}


\section{Introduction}

This paper belongs to a series devoted to the study of supposedly old
open clusters, which represent one of the best possibilities to 
study the chemical and dynamical evolution of our Galaxy. 
Open clusters provide unique information 
on the chemical abundances and gradients in the disc (e.g., 
Janes 1979, Panagia and Tosi 1981, Friel and Janes 1993); on the average 
stellar ages and radial velocities at different galactic radii (e.g., Janes 
and Phelps 1994, hereinafter JP94); 
and on the interactions between thin and thick discs (e.g.,
Sandage 1988). In addition, they are the only class of objects covering
a large range of distances (several kpc around the Sun) and
ages (from a few Myr to $\gsim$10 Gyr) and can, therefore, tightly constrain 
galactic evolution theories. Furthermore, in order to avoid misleading effects, 
it is mandatory to work with very accurate observational data and to treat them
homogeneously  (see Section 5, and e.g, Carraro and Chiosi 1994a,
hereinafter CC94, Friel 1995). 

In order to obtain this kind of homogeneity and to be able to study accurately
the metallicity and age distribution of open clusters with galactocentric
distance, we are analyzing systems of different ages and metallicities,
located at different galactic radii. 
To this end, we have obtained deep photometry for several clusters, 
and have supplemented these observations with published data of comparable quality to ensure 
a sample as large as possible of uniformly derived ages, metallicities and 
distance moduli. 

These quantities are derived from comparison of the observed colour-magnitude
diagrams (CMDs) to synthetic ones generated by a numerical code based on
stellar evolution tracks and taking into account theoretical and observational
uncertainties (Tosi et al. 1991). These simulations are much more powerful
than the classical isochrone fitting method to study the evolutionary status
of the analyzed region and have been successfully applied both to nearby
irregular galaxies (Marconi et al. 1995) and to galactic open clusters
(Bonifazi et al. 1990, Gozzoli et al. 1996, Bragaglia et al. 1997).

Our sample of old open clusters includes NGC2243 (age $\simeq$ 3 Gyr,
metallicity two tenths solar, Bonifazi et al. 1990), Collinder 261 (age
$\gsim$ 7 Gyr, almost solar metallicity, Gozzoli et al. 1996) and NGC6253 (age
$\gsim$ 3 Gyr, metallicity about twice solar, Bragaglia et al. 1997).

To them we add now NGC 2506, a moderately old open cluster already studied by
McClure et al. (1981, hereinafter MCTF) and Chiu \& van Altena (1981,
hereinafter CvA), located toward the galactic anticentre ($\alpha_{1950} = 7^h
57.6^m, \delta_{1950} = -10^{\circ}39^{\prime}$; l$_{\rm II}$ = 231$^{\circ}$,
b$_{\rm II}$ = $+10^{\circ}$). In section 2 we describe the observations and
data analysis; in Section 3 we present the derived CMDs involving U,B,G,V,R,I
photometry and discuss the presence of binary stars. In Section 4 we compare
observed and synthetic CMDs and derive metallicity, age, distance and
reddening. Finally, our findings will be discussed in Section 5.

\begin{figure*}
\vspace{14.5cm}
\caption{Map of the observed region of NGC 2506, from our photometry.
Pixel coordinates have been transformed to equatorial coordinates (year 2000)
making use of a set of stars identified on the Digitized Sky Survey.
The circle indicates the field observed by MCTF and has a diameter of 10 arcmin}
\includegraphics{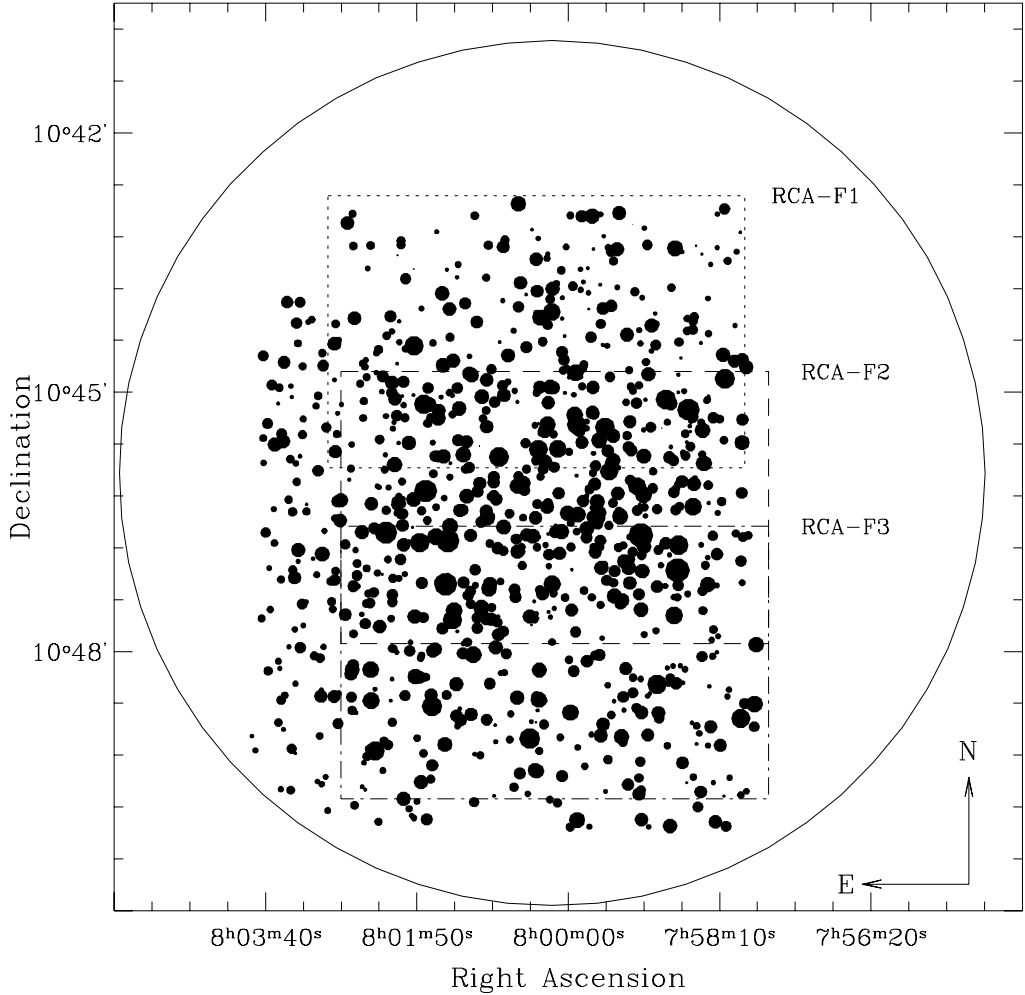}
\label{fig-map}
\end{figure*}

\section{Observation and data reductions}

Images of the cluster were obtained during two separate observing runs, in
1985 and 1993, with telescopes located on Cerro Tololo and Mt. Palomar
respectively (see Table 1). The total field covered is shown in
Fig.~\ref{fig-map}.

\subsection{Observations}

The first run was at the CTIO Blanco 4m telescope using the prime focus CCD
camera with RCA2.  The chip had good quantum efficiency even in the
near-UV/blue, but a large read-out noise (90 e$^-$ rms).  Using the doublet
corrector, this CCD at prime focus yielded a spatial sampling of 0.58 arcsec per
pixel.

Although this particular observing run consisted of several nights, the
cluster was observed for only two of these nights, both of which were
photometric. A minimum of 12 standard stars, as established by Graham (1982)
and by Landolt (1983), were observed several times each night and through a
range of airmass (1 to 2). The colours of the observed standard stars spanned
from \bv=$-$0.25 to \bv=1.35, encompassing the whole range of colours found in
the cluster members. The cluster observations were generally performed near
the meridian, and never with an airmass more than 1.2.

The filters used for the observations (UBGVRI) consisted of an unusual mixture
of passband definitions. The passbands of B and V were defined by standard
combinations of Schott coloured-filter glass. The passbands of R and I were
defined by interference filters, common to CTIO/KPNO at that time. The U and G
filters were those of Hamilton (unpublished). The U filter used in this case
was a combination of Schott UG5 with a liquid copper sulfate cut-off. Schott
UG5 is slightly redder and broader than the more standard UG2 commonly used at
CTIO since 1985. The G filter was specially designed to be a reasonable
representation of the passband defined by hypersensitized IIIaJ emulsions with
a two mm thick Schott GG385 filter, with two slight modifications. The cut-on
point was moved redward to 4150 \AA~ so as to clearly miss the influence of
the 4000 \AA~ break, and the cut-off point was moved slightly blueward so as
to reduce the contribution of bright night sky lines such as Hg 5460 \AA~ and
[OI] 5577 \AA~ on the overall detected sky background.  The measured passband
has a width of about 1000 \AA~ and a central wavelength of $\sim$ 4750 \AA.

As the field-of-view of this CCD was rather small by modern standards (about
$3' \times 5'$), the central part of the cluster was mapped with three fields,
each slightly overlapping with its neighbour (see Fig.~\ref{fig-map}).
Exposure times for the various frames varied from two seconds for the redder
passbands to 150 seconds for the ultraviolet exposures. At each pointing of
the telescope, several frames were taken in each filter.  These data were
later co-added, and it was this co-added frame that was analyzed. These short
exposures were necessary so that the bright giants of this cluster would not
exceed the dynamic range of the CCD.

\smallskip
The second observing run was conducted at the Palomar 60-inch telescope. 
The detector used was a thinned, back-side illuminated Tektronix
1024 x 1024 CCD at the Ritchey-Chr\'etien focus of this telescope. In this
case, the filters used were only B and V and these were defined 
similarly as those used during the first observing run.
The spatial sampling for this telescope/detector combination was 0.37 arcsec per
pixel (field of view about $6' \times 6'$).  The CCD data were not binned.
The exposure times were 300 seconds each for all frames, except for several of
those taken at the cluster centre, which were 100 seconds long.
Two series of 6 fields extending due northward and due eastward from the
cluster centre were taken, but here we only concentrate on the central frame.

\begin{table}
\begin{center}
\caption{Observing dates and filter sets}
\begin{tabular}{crcc}
\hline\hline
Date & Telescope, CCD & FOV & Filters\\
\hline
22-23/12/1985 & CTIO 4m, RCA      & $3'\times5'$ & UBGVRI\\
30-31/03/1993 & Palomar 60in, Tek & $6'\times6'$ & BV\\
\hline
\end{tabular}
\end{center}
\label{tab-obs}
\end{table}

\subsection{Reductions}

Basic CCD reduction of bias and dark current subtraction, trimming, flatfield
correction were applied.  Several frames of comparable seeing were co-added to
obtain the deepest frames in each band for the RCA data set. The co-added
frames as well as the single short exposures were analyzed. We applied the usual
procedure  for PSF study and fitting available in DAOPHOT--II (Stetson 1992) in
MIDAS  environment to all frames. The deepest V frame of each field has been
used to search for stellar objects, setting  the minimum photometric threshold
for object detection at 6$\sigma$ above the local sky background. All the
objects identified in V were then fitted in all the other bands.   The
internal errors were estimated by computing the $rms$ frame-to-frame scatter of
the magnitudes obtained for these stars, following the precepts of Ferraro et
al. (1990). The mean errors in B and V are shown in Table 2.

\begin{table}
\begin{center}
\caption{Mean errors for the magnitude bins}
\begin{tabular}{ccc}
\hline\hline
mag bin & Error B & Error V \\
\hline
$<$17.0  &0.004   &0.010\\
17.0-17.5  &0.010   &0.020\\
17.5-18.0  &0.013   &0.030\\
18.0-18.5  &0.017   &0.035\\
18.5-19.0  &0.020   &0.065  \\
19.0-19.5  &0.023   &0.095\\
19.5-20.0  &0.045   &0.150\\
\hline
\end{tabular}
\end{center}
\label{tab-err}
\end{table}

\subsection{Photometric calibrations}

Calibrations for the RCA data-set were carried out by a set of programs
written by Dr. Peter Stetson (unpublished; private communication 1985). Errors
in the fitted coefficients were generally less than 0.02 mag. The nightly
scatter was quite small for BGVR, and somewhat higher 
for U and I.  The coefficients for the various colour terms were from an average
over this particular observing run, as were the second-order 
extinction coefficients.  However, the zero-point and first order
extinction coefficients varied sufficiently from night-to-night that these were
determined separately for each night, assuming the aforementioned averaged
coefficients.
The resulting calibration equations are the following, where the lower case
symbols refer to the observed quantities, the upper case to the tabulated
magnitudes, and A stands for the airmass:

\[ u =U+1.470 -0.333  (U-B) + 0.410  A -0.075  (U-B)  A \]
\[~~~~~~{\rm Nightly~Scatter = 0.035 ~mag }\]
\[ b =B-0.069 +0.035  (B-V) + 0.310  A -0.090  (B-V)  A \]
\[~~~~~~{\rm  Nightly~Scatter = 0.007 ~mag }\]
\[ g =G-0.345 -0.091  (B-V) + 0.192  A +0.028  (B-V)  A \]
\[~~~~~~{\rm  Nightly~Scatter = 0.0035 ~mag}\]
\[ v =V+0.154 -0.042  (B-V) + 0.150  A +0.050  (B-V)  A \]
\[~~~~~~{\rm  Nightly~Scatter = 0.009 ~mag}\]
\[ r =R+0.13 + 0.12  A \]
\[~~~~~~{\rm  Nightly~Scatter = 0.018 ~mag}\]
\[ i =I+1.62 + 0.07  A \]
\[~~~~~~{\rm  Nightly~Scatter = 0.033 ~mag}\]

For the P60 data set, though taken under photometric conditions, 
no photometric standards were observed, as another program had priority.
These frames were calibrated from the photometric results of the first observing
run as there was sufficient overlap to be able to define the transformation
equations accurately.    

Final results of the reductions yielded magnitudes for 863 stars in the
composite sample. The table with UBGVRI magnitudes, coodinates,
and cross-identification with MCTF is available from the first author.

\begin{table}
\begin{center}
\caption{Completeness of our measurements in V and B, for the P60 field.
Each value is the average of 30 trials, where 10 \% of the total number 
of stars were added, following the luminosity function}
\begin{tabular}{c r r}
\hline\hline
Mag interv. & \%V &\%B \\
\hline
11.0-17.0 &  100   &100\\
17.0-17.5 &   98   &100\\
17.5-18.0 &   95   &100\\
18.0-18.5 &   92   &98\\
18.5-19.0 &   83   &94\\
19.0-19.5 &   51   &84\\
19.5-20.0 &   22   &58\\
20.0-20.5 &   5    &28\\
20.5-21.0 &   0    &11\\
\hline
\end{tabular}
\end{center}
\label{tab-compl}
\end{table}

\begin{figure*}
\vspace{8cm}
\caption{a) V, \bv~ colour-magnitude diagram for the total sample (P60+RCA); all
stars brighter than about the TO come only from the RCA chips. b) CMD taken
from MCTF; notice the much larger spread of the MS. c) CMD for the P60+RCA 
data set, but only for stars with membership probability $\ge$ 0.7.}
\includegraphics{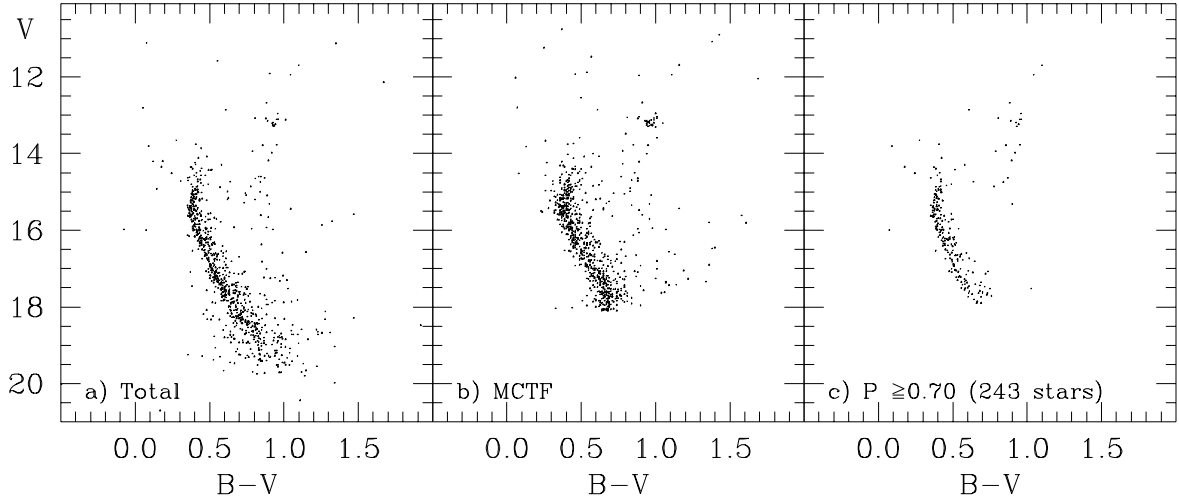}
\label{fig-cmd4}
\end{figure*}

\begin{figure*}
\vspace{14cm}
\caption{CMDs for the RCA sample alone: 
a) $V,B-V$; b) $V,G-V$, c) $V,V-R$; d) $V,V-I$; e) $U,U-B$;
f) $U,U-V$; g) $U,U-R$; h) $U,U-I$}
\includegraphics{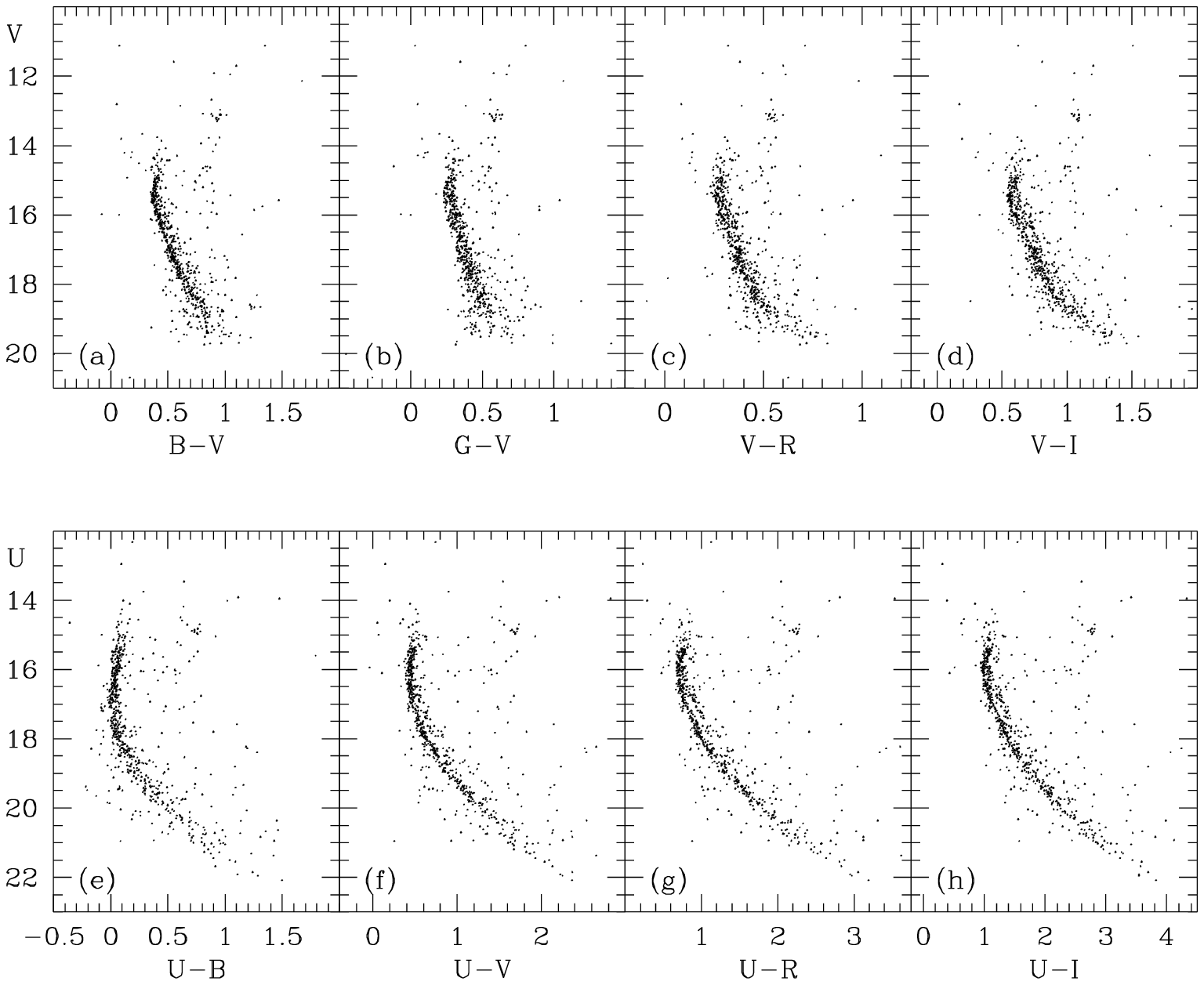}
\label{fig-cmd5}
\end{figure*}

\subsection{Completeness analysis}

We tested our luminosity function in the V and B band for completeness, using
the routine ADDSTAR in DAOPHOT--II. In short, we added to  the original V and
B frames a pattern of artificial stars ($\simeq$ $10\%$ of the total in each
magnitude bin), at random positions. The frames were reduced
in the same way described above. We considered as ``recovered'' only those
stars found in their given position and magnitude bin. The completeness was
then derived as the ratio $N_{recovered}/N_{added}$ of the artificial stars
generated. We performed 30 trials for band and the final averaged results are
reported in Table 3.

\section{The colour-magnitude diagrams}

We present the colour-magnitude diagram (CMD) obtained from our reductions 
in Figs~\ref{fig-cmd4} and ~\ref{fig-cmd5}.
Figs~\ref{fig-cmd4} refers to our total sample (P60+RCA frames), and is thus
limited to the B and V filters. 
Fig.~\ref{fig-cmd5} instead contains only stars found in the three RCA fields,
and diagrams are shown for various combinations of the UBGVRI filters.

\begin{figure}
\vspace{10cm}
\caption{Comparison between MCTF's and our photometry in B and V for the 462
stars in common (see Sect. 3.2)}
\includegraphics{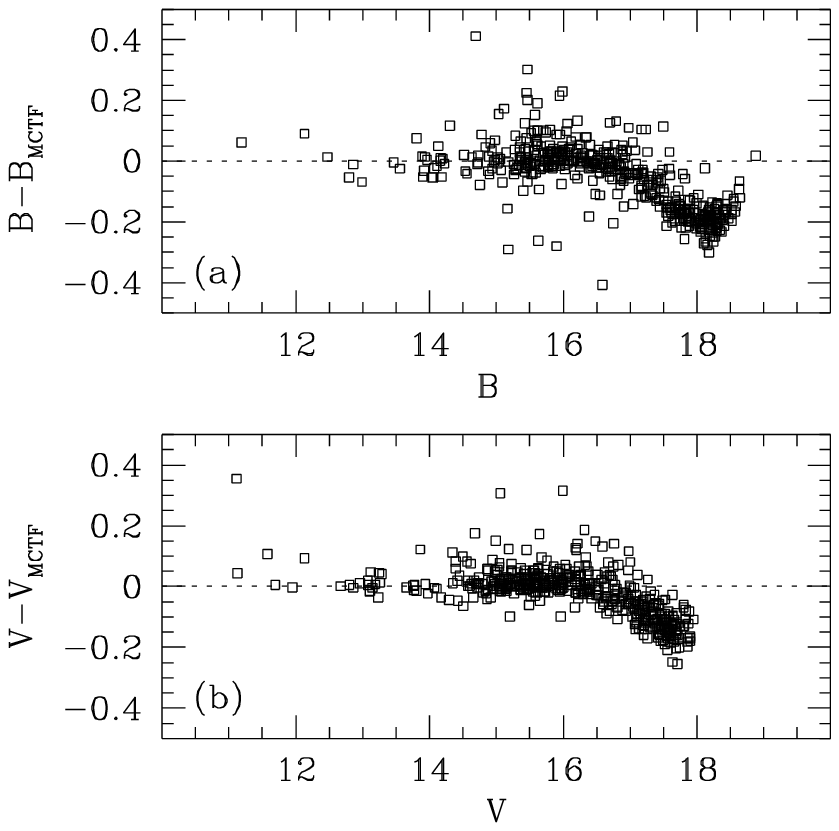}
\label{fig-com}
\end{figure}

Our $V,B-V$ diagram (Fig.~\ref{fig-cmd4}(a)) agrees reasonably well with the
one obtained by MCTF (Fig.~\ref{fig-cmd4}(b)), but is much better defined.
MCTF's photographic study covered a diameter of 10 arcmin centred on the
cluster, and our field is mostly contained in theirs zones I and II (see fig.
1 in MCTF, and our Fig.~\ref{fig-map}). In Fig.~\ref{fig-com} we show the
differences between our magnitudes and theirs. 
These differences are less than 0.1 mag down to B or V = 17. 
However, there is a clear trend fainter than this, in the sense of our measures 
being $\lsim$ 0.2 mag brighter than those of MCTF. 

Our Turn-Off (TO) point, defined as the bluest point of the top of the 
main sequence, is located at $V=14.7$, $B-V=0.35$. 
TO morphology\footnote[1]{Proper interpretation of the 
TO morphology is complicated by the combination
of the normal main sequence (MS) defined by single-stars, and by the binary
sequence, which is 0.75 mag brighter that the normal MS.
This binary MS is apparent in all the CMDs.}
shows a definite ``hook'', similar e.g. to the one in NGC6253 (Bragaglia et
al. 1997), typical of an intermediate old age. The subgiant and red giant
branches (SGB, RGB) are extended and well delineated.  The red clump, which
represents the locus of core-He burning stars, is found at $V=13.2$,
$B-V=0.92$. There also are objects above the RGB, probably asymptotic branch
stars; of them, at least 2 are cluster members (CvA, MCTF). At odds with many
open clusters, NGC2506 does not seem to have a noticeable number of Blue
Straggler Stars (see especially Fig.~\ref{fig-cmd4}(c), where only probable
cluster members are plotted).

Given the magnitudes of the TO and red clump, we derive 
$\delta$V=$V_{TO}-V_{clump}$=1.5
which, following JP94 relation for the Morphological Age 
Index (MAI), suggests an age of 2.5 Gyr. We
will see in section 4 that our derived age is actually younger.

\subsection{The Lower Main Sequence}

Our MS extends to about 2 mag fainter than that presented by MCTF.
There appears to be a discrepancy with their diagram, as we see no 
blueward migration of the MS at these lower luminosities.
We attribute such a peculiarity to non-linear effects in 
their photographic material near the plate limit.

Our cluster MS remains well defined also towards the faint end, though 
apparently increasingly less populated towards fainter luminosities. 
A similar effect had already been noticed by Scalo (1986), who used 
Chiu \& van Altena's data, and is not uncommon among open clusters. In this
case, though, it seems to be perfectly explained by incompleteness in
photometry at the faint limit (see Section 4 and Fig.~\ref{fig-sim}). In other
words, there is no indication of a flattening of the present-day 
mass function at the low-mass end, as seen in many other clusters 
(e.g. NGC6253, Bragaglia et al. 1997 or, as extreme case, NGC 3680, 
Nordstrom et al. 1996) and usually ascribed to dynamical evolution. 
This may be due to the Galactic position of the cluster.
According to Carraro \&
Chiosi (1994b), who determined the orbit of this cluster using radial
velocity information, NGC2506 has never moved very far away from its
birthplace, having an orbit confined in distance from the Galactic centre
between 10.7 and 11.6 kpc, and which doesn't extend beyond 0.6 kpc
from the Galactic plane. Thus, NGC2506 has probably suffered less than other
clusters from
encounters with giant gas clouds which produce stripping of cluster members.

\subsection{Field contamination}

\begin{figure*}
\vspace{8cm}
\caption{CMD for the MS, derived from P60 fields, at increasing distance from
the cluster centre, shown in panel a). The other CMDs are for areas
distant respectively: b) 6 arcmin, c) 9 arcmin, d) and e) 31 arcmin}
\includegraphics{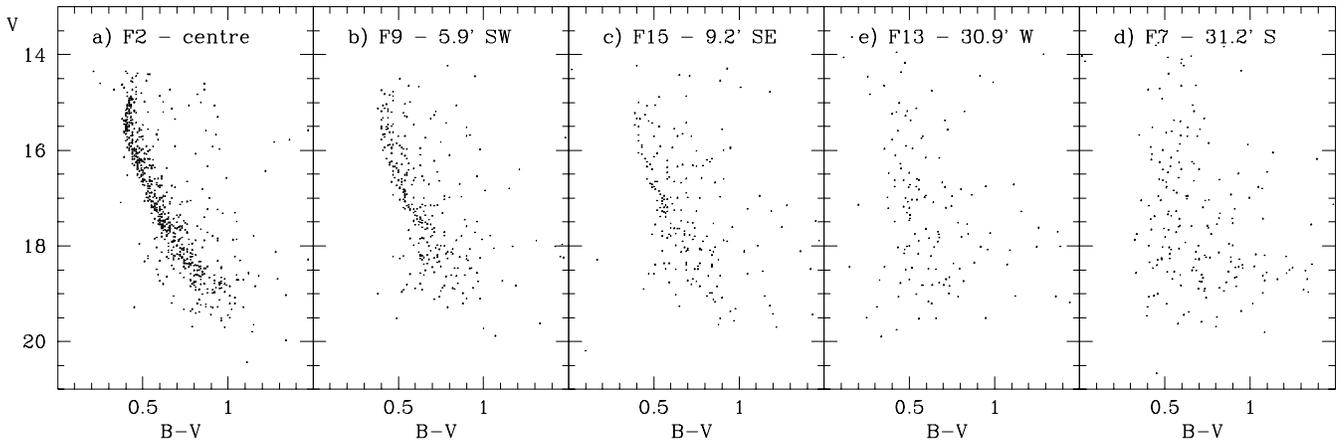}
\label{fig-rad}
\end{figure*}

Field contamination is likely to be small in the anti-centre direction, but
separation of field and cluster stars simply on the basis of photometry is
neither easy nor unambiguous, since the cluster extends quite far (see
Fig.~\ref{fig-rad} where panel a) corresponds to the cluster centre, and
panels b) to e) to increasing distances). Even at about 30 arcmin from the
centre there might be a slight cluster component mixed with the field stars.
We will not discuss the cluster dimension and radial distribution, since these
will be amply treated by Testa \& Hamilton (1997).

Chiu \& van Altena (1981) published a membership study of this cluster for the
stars in MCTF; their table 2 contains 801 stars down to V about 18. Of the 863
stars in our catalogue, 462 were cross-identified with MCTF on the basis of
position and magnitude, while 401, mostly faint, are present only in our
photometry.
The synthetic CMDs described in section 4 will be compared to our whole sample,
since we lack membership information for V=18--20 and there is no way to assess
the incompleteness of the membership study at the faint limit.
The CvA data are instead very useful at the bright end, since we can 
be sure of how the MS behaves near the TO, of the position of the subgiant and
red giant branches, and of the location of the clump (see
Fig.~\ref{fig-cmd4}(c), where we show only those stars with membership
probability $\ge$ 0.70).

\begin{figure*}
\vspace{7cm}
\caption{Left panel: $U,U-I$ CMD for stars on/near the MS in the RCA-F1
field; the MS ridge line and a secondary sequence 0.7 mag above it (representing
the equal-mass binary systems) are also shown. Middle panel: histograms of
the distance in colour from the MS ridge line of all stars, divided in 4
magnitude intervals; the arrows indicate roughly the position of the
equal-mass binary sequence in the middle of each mag bin, 
as measured from the left panel. Right panel: histogram of the magnitude
distance from the MS ridge line; the arrow indicates the position of stars
0.7 mag brighter than this line.} 
\includegraphics{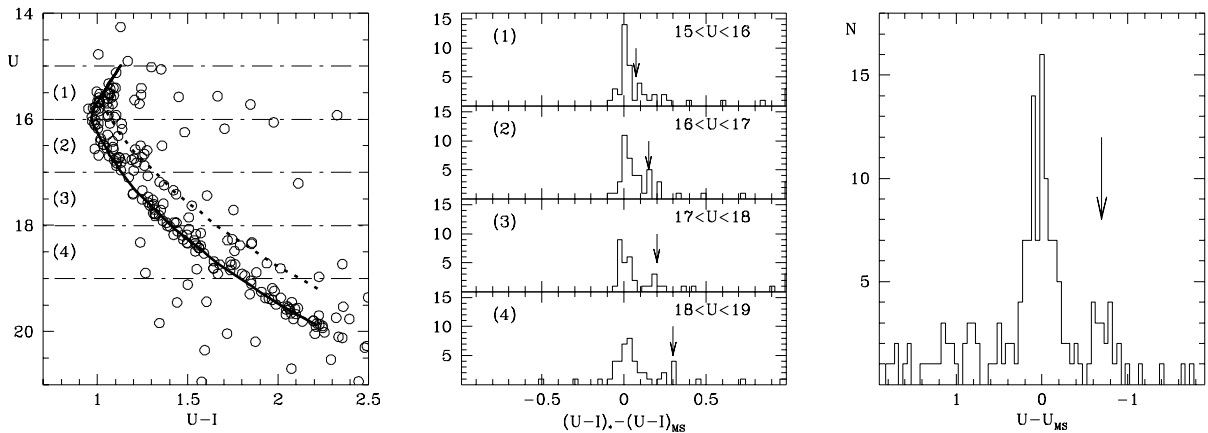}
\label{fig-bin}
\end{figure*}

\subsection{Binary stars}

Most open clusters show indications of a sizeable binary population, detected
either spectroscopically (see e.g. Mermilliod \& Mayor 1989, 1990) or
photometrically (see e.g.  M 67, Fan et al. 1996, or two clusters quite similar
to NGC2506: NGC 2243, Bonifazi et al. 1990, and NGC 2420, Anthony-Twarog et al.
1990). 

NGC2506 has been unsuccessfully surveyed for close/moderately close binaries in
the past: Cameron \& Reid (1987) found no candidate binary among 26
subgiant/lower red giant cluster members. Their technique was based on the
detection of chromospheric emission in the CaII H+K lines due to enhanced axial
rotation produced by orbital locking and was claimed to be sensible to periods
of about 1 to 50 days. Kaluzny \& Shara (1988) found no contact binaries of the
W UMa type, with periods in the range a few hours -- 1.5 days, in their CCD
survey of 6 old open clusters, among which was NGC2506. 
MCTF, on the basis of the scatter of the MS, estimated a crude 50~\% of
binaries in NGC2506.

The $V,B-V$ diagram is not the most suitable to separate the secondary sequence
due to binary systems from the single-star MS: much better resolution in
colour is obtained in the CMD's involving the U band (see Fig.~\ref{fig-cmd5}).
To quantify the visual impression of well populated binary star sequences, we
have performed a simple experiment on RCA-F1, our best field both for
photometric accuracy and crowding conditions, using the $U,U-I$ CMD. After
defining a MS ridge line, we have measured the distance in colour of every star
from this MS and plotted the distance histograms in four separate magnitude
bins (see Fig.~\ref{fig-bin} -- left panel -- for a definition of the MS and
magnitude intervals, and Fig.~\ref{fig-bin} -- middle panel -- for the
histograms). A secondary peak in the colour distribution is clearly present:
counting stars in it and in the ``MS peak'', we derive a binary frequency of
about 17~\%. Finally, the right panel of Fig.~\ref{fig-bin} shows the 
histogram of the distance of each star in RCA-F1 from the MS ridge line: here
too the secondary peak is clearly prominent.

This value is referred to the central part of the cluster, but since NGC2506
does not suffer from segregation of the more massive objects toward the centre,
as seen from the luminosity function (Fig.~\ref{fig-sim}), we may assume it is
representative of the cluster as a whole.

\section{CLUSTER PARAMETERS}

In the case of NGC2506, there are estimates within the literature of the age,
reddening and metallicity, with the latter two being fairly reliable: MCTF
derived \ebv=0.05 from photoelectric measures and Friel \& Janes (1993)
obtained [Fe/H]=$-0.52 \pm 0.07$ from medium-resolution spectroscopy of 5
cluster members.

To derive our own values for these parameters, we have applied to NGC2506 the
approach of CMD simulations described by Tosi et al. (1991). This technique
has been employed for three other old open clusters (Bonifazi et al. 1990,
Gozzoli et al. 1996, Bragaglia et al. 1997). Due to the existing estimate of
the cluster metal content, we have restricted the sample of stellar evolution
tracks adopted to construct the synthetic CMDs to those sets with
metallicities near that value. The major features of these sets are summarized
in Table 4 where their origin is also given. For each set of stellar models,
we have performed several MonteCarlo simulations for any reasonable
combination of age, reddening and distance modulus.

\begin{table*}
\begin{center}
\caption{Stellar evolutionary tracks adopted for the synthetic CMDs}
\begin{tabular}{lccccll}
\hline\hline
Model & Y & Z & M$_{min}$ (M$_{\odot}$) & M$_{max}$ (M$_{\odot}$) 
      & Reference & Notes\\
\hline
FRANEC & 0.27 & 0.01 & 0.6 & 9 & Castellani et al. 1993 & to AGB-tip, LAOL op.\\
FRANEC & 0.27 & 0.02 & 0.6 & 9 & Castellani et al. 1993 & to AGB-top, LAOL op.\\
Geneva & 0.26 & 0.008 & 0.8 & 120 & Schaerer et al. 1993 & only to RGB-tip\\ 
Geneva & 0.30 & 0.02  & 0.8 & 120 & Schaller et al. 1992 &Charbonnel et 
                                                         al. 1996 for e-AGB \\
Padova & 0.28 & 0.008 & 0.6 & 120 & Alongi et al. 1993  & to AGB-tip\\
Padova & 0.28 & 0.02  & 0.6 & 120 & Bressan et al. 1994 & to AGB-tip\\
\hline
\end{tabular}
\end{center}
\label{tab-mod}
\end{table*}

\begin{figure*}
\vspace{12cm}
\caption{Models in better agreement with the data: the synthetic CMDs (to be
compared with that in Fig. 3(a)) are in the top panels and the corresponding
luminosity functions in the bottom panels (lines for the models and open
circles for the data). Panels (a) and (d) correspond to FRANEC with Z=0.01,
panels (b) and (e) to Padova with Z=0.02, and panels (c) and (f) to Geneva
with Z=0.008. See text for details.}
\includegraphics{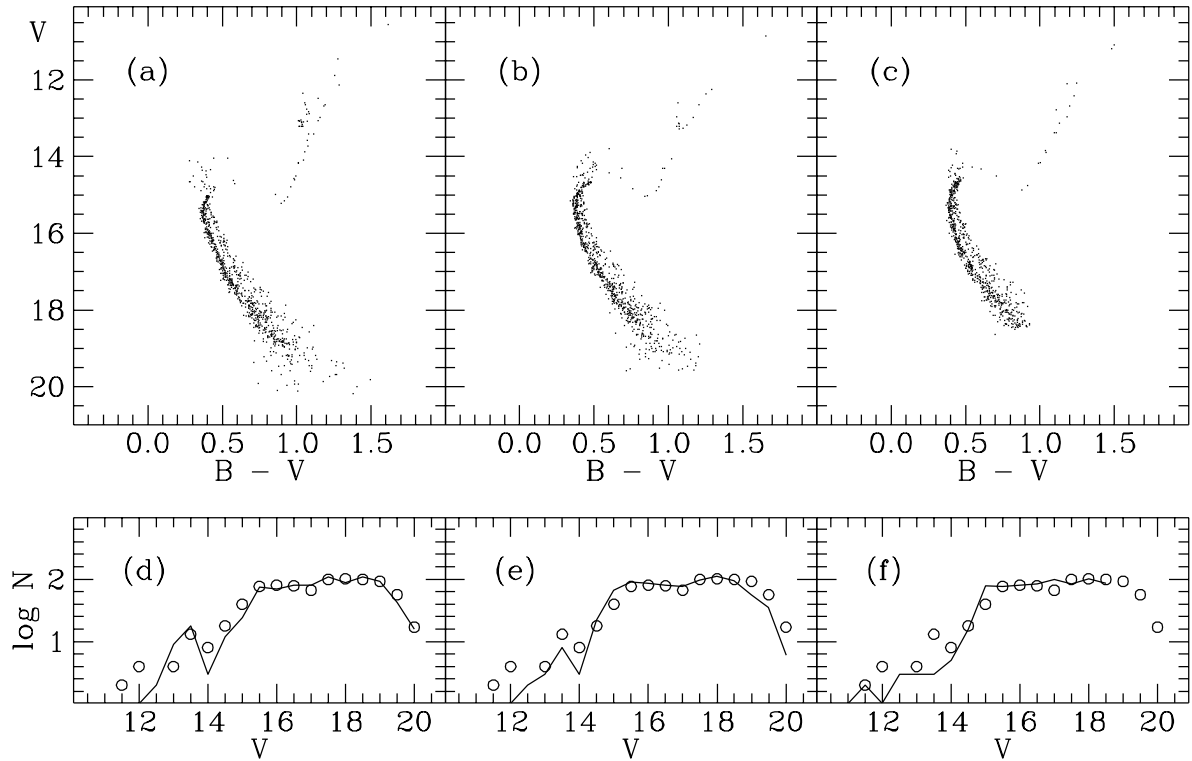}
\label{fig-sim}
\end{figure*}

The incompleteness factors and the photometric errors in each magnitude bin
assigned to the synthetic stars in each photometric band are those derived
from the observed data and mentioned in section 2. Naturally, the number of
stars in the synthetic diagram is the same as that in the empirical CMD of
Fig.~\ref{fig-cmd4}(a). To compare the model predictions with the
observational distributions (both CMD and luminosity function, LF), we have
taken into account the number of probable non members, estimated with the help
of CvA's diagrams with different membership probability. However, we
have not removed any of the stars above the MS turn-off, and this is why the
cluster LF (represented by the open circles in the bottom panel of
Fig.~\ref{fig-sim}) is always slightly higher than the curve predicted by the
models in the brightest magnitude bins.

Since the data show evidence for a significant fraction of binary stars among
the cluster members, we have taken them into account in the synthetic diagram.
A mass ratio has been associated to each system via random extractions from a
flat distribution. We have then followed the prescriptions given by Maeder
(1974) to attribute colours and magnitudes to systems with different
primary/secondary mass ratios (see Bragaglia et al. 1997 for more details).
It is interesting to notice that all the stellar models lead to synthetic
CMDs in better agreement with the data when the assumed fraction of binaries is
30$\%$. As already discussed by Maeder (1974) and Fan et al. (1996), this 
percentage is larger that that derived by counting the stars on the right
side of the MS (see section 3.3) and the difference is due to the combination
of several selection effects.

We wish to note here that $all$ of the simulated cases yield red-giant branches
that are slightly redder than the observed ones and that in all cases, except
in the FRANEC models with Z=0.02, this colour excess is already visible at the
base of the RGB. We ascribe this effect to the
uncertainty in the photometric conversions from the theoretical plane
(luminosity and effective temperature) to the observational one (magnitude and
colour):  the temperature-colour conversion is the one that is most affected.
The fact that, at the same colours, the discrepancy is not observed in the
lower MS (except perhaps in the Padova models with Z=0.02) suggests that the
inadequacy of the temperature-colour conversion does not concern all the
cool stars, but only those in  low-gravity conditions, a circumstance foreseen
by Bessel et al. (1989).

\subsection{Results with FRANEC stellar models}

Both of the examined sets of FRANEC evolutionary tracks provide synthetic CMDs
and LFs in agreement with the observed data, although with slight differences
in the choices of the parameters.
With the set at Z=0.01, the best fit is obtained assuming an age of 1.6 Gyr,
a reddening \ebv=0.07 and a distance modulus \mmm=12.6. As shown in panels
(a) and (d) of Fig.~\ref{fig-sim} these stellar models reproduce quite
well all the observed features of the cluster (namely: the magnitude and colour
distribution of the stars, the relative number of stars in the various
evolutionary phases and their morphology, including the MS gaps). 

A very good agreement with the data, practically undistinguishable from that
shown in Fig.~\ref{fig-sim} (a) and (d), is achieved also with the FRANEC
set with Z=0.02. In this case, the best model indicates roughly the same age 
(1.5 Gyr) and distance modulus (\mmm=12.7), but a lower reddening (\ebv=0.01), 
to compensate the intrinsically redder colours corresponding to the doubled
metallicity.
These tracks have a metal content nominally larger than that spectroscopically
attributed to NGC2506. However, according to their authors, the FRANEC 
tracks (i.e., both the set with Z=0.01 and with Z=0.02) actually correspond to 
models with metallicity about half of their nominal value, once the effect of 
using the old LAOL opacities rather than the most recent OPAL opacities is 
taken into account. Both sets are, therefore, roughly compatible with the 
spectroscopic metallicity [Fe/H]=$-$0.5 derived by Friel \& Janes (1993).

\subsection{Results with Padova stellar models}

The stellar evolutionary models computed by the Padova group are slightly
brighter than the corresponding FRANEC models, probably because they take into
account the possible overshooting from the convective core, and a larger core
implies higher brightness. For this reason, the ages derived from overshooting
models are usually older than those derived with standard treatment
of the convective regions. The luminosity difference between FRANEC and
Padova models can be appreciated comparing panels (a) and (b) of 
Fig.~\ref{fig-sim}: in both cases the faintest stars have mass 0.6 M$_{\odot}$,
but they appear brighter in the Padova than in the FRANEC diagram.
Notice also the rounder shape of the MS turn-off region of the Padova CMDs.

In the case of NGC2506, the age difference is, however, very small.
With the Padova tracks with Z=0.008, the best fit to the cluster features
is obtained for an age 1.7 Gyr and assuming \ebv=0.05 and \mmm=12.5, 
confirming that the literature reddening is the most appropriate when the
cluster is assigned the spectroscopic metallicity. 

It must be emphasized, however, that an equally good fit is attained with
the Padova solar metallicity models, in which case the reddening must be
smaller to compensate for the intrinsic redder colour of more metal rich
stars. The best case for Z=0.02 assumes \ebv=0.0, \mmm=12.6, an age of 1.7 Gyr,
and is shown in Fig.~\ref{fig-sim} (b) and (e). Notice that the metallicity
significantly affects only the reddening determination.

These results are in striking agreement with those derived from the FRANEC
models described above. This shows that in the mass range of
the stars in NGC2506 (stars at the MS turn-off have masses  1.7-1.9
M$_{\odot}$) the overshooting effect on the age of the stellar models
is moderate.

\subsection{Results with GENEVA stellar models}

The Geneva stellar models take into account the overshooting off convective
cores, although with a formalism different from that applied in the Padova
tracks. For a given stellar mass, these are the brightest models of the
various sets examined here and should provide the oldest age. 

Figs~\ref{fig-sim} (c) and (f) show the Geneva model in better agreement 
with both the CMD and the LF of NGC2506: with Z=0.008, it indicates an age
of 2.2 Gyr and requires \ebv=0.05 and \mmm=12.5. The helium burning phases
have not been computed for the Geneva stellar tracks with this metallicity.
They provide a lower number of post-MS stars than the FRANEC and 
Padova models and the clump is not visible. This
leads to a worse reproduction of the empirical LF in the magnitude bins
brighter than V$\simeq$14. The absence of the predicted curve
in the faintest magnitude bins (Fig.~\ref{fig-sim}(f)) is due to the fact
that the Geneva stellar evolutionary tracks have been computed only for
stars more massive than 0.8 M$_{\odot}$ and should not be taken as a model
failure. 

The Geneva tracks with solar metallicity have instead been computed also
for the later evolutionary phases and show the clump in the right CMD
position. However, they provide in general a worse reproduction of the 
cluster features, probably because of their excessive metallicity. 
The derived parameters are, however, consistent with the others: the age is 
in the range 1.7-2.0 Gyr, \ebv=0.0 and \mmm=12.5 or 12.6.

\section{Summary and Discussion}

\begin{table}
\begin{center}
\caption{Summary of best-fits results for distance modulus, age, and reddening 
for all models used}
\begin{tabular}{lrccc}
\hline\hline
\multicolumn{1}{c}{Model} &\multicolumn{1}{c}{Z}
&\multicolumn{1}{c}{(m-M)$_0$}
&\multicolumn{1}{c}{$\tau$ (Gyr)} &\multicolumn{1}{c}{\ebv} \\
\hline
FRANEC & 0.01  &12.6 & 1.6 & 0.07 \\
FRANEC & 0.02  &12.7 & 1.5 & 0.01 \\
Padova & 0.008 &12.5 & 1.7 & 0.05 \\
Padova & 0.02  &12.6 & 1.7 & 0.00 \\
Geneva & 0.008 &12.5 & 2.2 & 0.05 \\
Geneva & 0.02  &12.5-12.6 & 1.7-2.0 & 0.00\\
\hline
\end{tabular}
\end{center}
\label{tab-ris}
\end{table}

We have determined for NGC2506 a confidence interval for distance, age,
reddening and metallicity (see Table 5).

a) Distance: in all our trials with the six sets of models we have a
best fit for (m-M)$_0$ in the range 12.5 -- 12.7. This is to be compared with
the 12.2 value found by MCTF using the Ciardullo \& Demarque (1977) isochrones.
We must emphasize that with none of our models are we able 
to provide a stellar distribution in the CMD compatible with the data
with such a low distance modulus. CC94 have derived (m-M)$_0$=12.5, which
is in agreement with our derivation.

b) Age: it varies from 1.5 to 2.2 Gyr depending on the adopted set of
stellar evolutionary models, with better fits for ages 1.5-1.7 Gyr.
This value is much smaller than the 3.4 Gyr derived by MCTF and the 6 Gyr 
found by Barbaro \& Pigatto (1984). Only part of the age overestimates
of those studies is attributable to their adopted shorter distance, which 
naturally implies lower intrinsic brightness and, therefore, older age. Most of
the problem resides in the use of the isochrone fitting method with 
older, less accurate, stellar models combined with the uncertainty in the
definition of the MS turn-off (fig.6 in MCTF clearly shows that what we have
considered as the turn-off is closer to their 2 Gyr than to the 3 Gyr 
isochrone).
Our age is also smaller than the 2.5 Gyr cited in Friel (1995), and based on
the MAI defined by JP94: this point will be shortly discussed later. It is
instead in excellent agreement with the 1.9 Gyr found by CC94, who used the
same distance modulus and Padova tracks adopted here.

c) Reddening: it strongly depends on the metallicity of the models, but
varies from 0 to 0.07, in good agreement with the 0.05 value found by MCTF. 

d) Metallicity: we have obtained excellent fits using the sets of models with
metal content lower than solar, in agreement with the spectroscopic value.
However, in one case (Padova models), we have also reached satisfying results 
with the solar metallicity set. This is because a tight determination of
the metallicity cannot be done on the basis of the
photometry alone. While spectroscopic values are to be preferred in general,
we remind that the [Fe/H]=--0.52 found by Friel \& Janes (1993) is based on
medium-low resolution spectra, calibrated against a metallicity scale. High
resolution direct spectroscopy will be able to put a more stringent limit on
metallicity, and our $V-I$ values could be used to derive input temperatures
for the model fitting analysis.

As a by-product of our theoretical simulations, we can also provide a lower
limit to the initial mass of the cluster within the examined region; that
is the minimum mass of the stars that must have formed there to account for the
current stellar content. This value turns out to be 
$<$M$_i>\simeq2.6\times10^{3}$M$_{\odot}$ 
(ranging from 1.6 to 3.0 depending on the adopted evolutionary
tracks) and is a lower limit to the actual initial mass of the cluster region,
since it does not take into account the possible effect of subsequent stellar
evaporation.

The age differences presented at point b) show, in our opinion, how much safer 
it is to date the clusters with the synthetic CMDs rather than with isochrone
fitting. Both methods provide results dependent on the adopted stellar
evolution models. However, the synthetic CMDs, on the one hand, are
independent of any other parameter and, on the other, are constrained by
the many morphological features (e.g., shape of the MS, SG and RG branches,
position of the clump, of the TO, of the MS gaps), by the stellar
distributions with magnitude and colour in the diagram, and by the proportion
between stars in different evolutionary phases. 
Isochrone dating is strongly affected by the 
choice of the MS turn-off (which can be
quite subjective) and has no further constraint aside from the average stellar
distribution in the CMD. Besides, it cannot account for the colour and
magnitude spread due to photometric errors or for the effect of incompleteness
on the relative number of stars in the various phases.

Our study of open clusters is aimed at better understanding the chemical
and dynamical evolution of the Galaxy; as stated in the Introduction, their ages
(and distances) are among the soundest derived for the disc population. 
Nevertheless, homogeneity in dating them is fundamental if one wishes to 
preserve a meaningful age ranking.

In this respect, the effort made by CC94, who re-analyzed ten well observed 
clusters  and defined a relation between a measurable index
($\Delta$V) and age, is very important. Carraro \& Chiosi defined $\Delta$V 
as the difference in magnitude between the TO and the red clump (like the 
$\delta$V used by JP94), but taking into account the possible
misplacement of the TO due to the binary sequence by adding an extra 0.25 mag
to the value found from the CMD. They have then derived the age of
the ten clusters with the synthetic CMD method and obtained a linear fit
between Log(age) and $\Delta$V. They have then applied this relation to derive 
ages for a sample of 26 clusters with reliable photometry published. 

JP94 used a similar approach. They used literature values for ages (stressing
the fact that in doing so they lacked the homogeneity one would have wished
for), and chose the 7 more reliable open clusters. They derived a non-linear
relation between $\delta$V and Log(age), using also values for Globular
clusters. In this way their relation is quite sound for high $\delta$V values,
and furthermore, these are intrinsically simpler to measure, since for older
ages the red clump is more populated, hence easier to locate precisely. At
intermediate ages, like in the case of NGC2506, there seems instead to be a
discrepancy (see Table 6), with ages from $\delta$V too high.
The same problems exists for the CC94 relation (their eq. 3): their measured
$\Delta$V=1.75 for NGC2506 yields an age of about 2.5 Gyr, which decreases to
1.8 Gyr using instead 1.5 (our value).

Comparing CC94's $\Delta$V and JP94's $\delta$V for the 26 clusters common to 
CC94 and JP94, one finds a mean difference of about 0.15 mag, in the sense of 
larger values for CC94. Restricting the comparison to the 6 clusters in common 
chosen among the 7 best clusters used by JP94 to build the MAI relation yields 
a mean difference of 0.25 mag (excluding the case of NGC3680, a scarcely
populated cluster, for which the difference is 0.7 mag). For these 6 objects
the mean age difference is about 0.1 Gyr.
We then conclude with this $caveat$ about the MAI age ranking
system, which could lead to conspicuous errors in individual ages. The effort
to build a more refined relation is huge, but worth pursuing.

\begin{table}
\begin{center}
\caption{Comparison of ages derived for our sample of old clusters, from MAI
(JP94) and by CC94}
\begin{tabular}{lccccc}
\hline\hline
Cluster & $\tau$ & $\delta$V & MAI  &$\Delta$V & CC94  \\
        & (Gyr)  &  (mag)    &(Gyr) &  (mag)   & (Gyr)  \\
\hline
NGC2506 &1.5-2.2 & 1.5 & 2.5 & 1.75 & 1.9 \\
NGC6253 & 3.0    & 2.0 & 4.4 &      &     \\
NGC2243 & 3-5    & 2.2 & 5.6 & 2.15 & 4.5 \\
Cr261   & 7-11   & 2.6 & 9.5 &      &     \\
\hline
\end{tabular}
\label{tab-mai}
\end{center}
\end{table}

\bigskip\bigskip\noindent
ACKNOWLEDGEMENTS

The bulk of the numerical code for CMD simulations has been provided by Laura
Greggio. The FRANEC, Geneva and Padova evolutionary tracks were kindly made
available by their authors. The authors would like to thank Dr. Neill Reid for
a critical reading of the manuscript and Dr. Vincenzo Testa for help with
various aspects of the photometric reductions. We made use of the digitization
of photographic data obtained at the UK Schmidt Telescope (Digitized Sky
Survey, produced at the Space Telescope Science Institute under US Government
grant NAG W-21). This research has made use of the Simbad database, operated
at CDS, Strasbourg, France.


\begin{thebibliography}{}

\bibitem{} Alongi, M., Bertelli, G., Bressan, A., Chiosi, C., Fagotto, F., 
 Greggio, L., Nasi, E. 1993, AAS, 97, 851

\bibitem{} Anthony-Twarog, B.J., Kaluzny, J., Shara, M.M., Twarog, B.A. 1990,
 AJ, 99, 1504

\bibitem{} Barbaro, G., Pigatto, L. 1984, A\&A, 136, 355

\bibitem{} Bessel, M.S., Brett, J.M., Scholz, M., Wood, P.R. 1989, AAS, 77, 1

\bibitem{} Bonifazi, A., Fusi Pecci, F., Romeo, G., Tosi, M. 1990, 
 MNRAS, 245, 15
 
\bibitem{} Bragaglia, A., Tessicini, G., Tosi, M., Marconi, G., Munari, U.
 1997, MNRAS, 284, 477

\bibitem{} Bressan, A., Fagotto, F., Bertelli, G., Chiosi, C. 1994, AAS 100, 647

\bibitem{} Cameron, A.C., Reid, N. 1987, MNRAS, 224, 821

\bibitem{} Carraro, G., Chiosi, C. 1994a, AA, 287, 761, CC94

\bibitem{} Carraro, G., Chiosi, C. 1994b, AA, 288, 751

\bibitem{} Castellani, V., Chieffi, A. \& Straniero, O. 1993, ApJS 78, 517

\bibitem{} Charbonnel, C., Meynet, G., Maeder, A., Schaerer, D. 1996, A\&AS
  115, 339

\bibitem{} Chiu, L.-T.G., van Altena, W.F. 1981, ApJ, 243, 827

\bibitem{} Ciardullo, R.D., Demarque, P. 1977, Trans. Yale Univ. Obs, 33, 1

\bibitem{} Fan, X., et al 1996, AJ, 112, 628

\bibitem{} Ferraro, F.R., Clementini, G., Fusi Pecci, F., Buonanno, R.,
 Alcaino, G. 1990, A\&AS, 84, 59

\bibitem{} Friel, E.D. 1995, ARAA, 33, 381

\bibitem{} Friel, E.D., Janes, K.A. 1993, AA, 267, 75

\bibitem{} Gozzoli, E., Tosi, M., Marconi, G., Bragaglia, A. 1996, MNRAS,
 283, 66

\bibitem{} Graham, J.A. 1982, PASP, 94, 244

\bibitem{} Janes, K.A. 1979, ApJS, 39, 135

\bibitem{} Janes, K.A., Phelps, R.L. 1994, AJ, 108, 1773, JP94

\bibitem{} Kaluzny, J., Shara, M. 1988, AJ, 95, 785

\bibitem{} Landolt, A.U. 1983, AJ, 88, 439 


\bibitem{} Maeder, A., 1974, A\&A, 32, 177

\bibitem{} Marconi, G., Tosi, M., Greggio, L., Focardi, P. 1995, AJ, 109, 173

\bibitem{} McClure, R.D., Twarog, B.A., Forrester, W.T. 1981, ApJ, 243, 841

\bibitem{} Mermilliod, J.C., Mayor, M. 1989, A\&A, 219, 125

\bibitem{} Mermilliod, J.C., Mayor, M. 1990, A\&A, 237, 61

\bibitem{} Nordstrom, B., Andersen, J., Andersen, M.I, 1996, A\&AS, 118, 407

\bibitem{} Panagia N., Tosi M. 1981, A\&A,  96, 306

\bibitem{} Sandage, A.R. 1988, A.G.D. Philip, L. Davis, eds, Calibration of 
 stellar ages, (Schenectady USA), p.43

\bibitem{} Scalo, J.M. 1986, Fund.Cosm.Phys. 11, 1

\bibitem{} Schaerer, D., Meynet, G., Maeder, A., Schaller, G.
  1993, A\&AS 98, 523

\bibitem{} Schaller, G., Schaerer, D., Meynet, G., Maeder, A., 
  1992, A\&AS 96, 269

\bibitem{} Stetson, P.B. 1992, User's Manual for DAOPHOT-II

\bibitem{} Testa, V., Hamilton, D. 1997, in preparation

\bibitem{} Tosi, M., Greggio, L., Marconi, G., Focardi, P. 1991, AJ, 102, 951

\end{thebibliography}
\end{document}